\documentclass[doublecol]{epl2} 

\usepackage{amsmath}
\usepackage{amssymb}
\usepackage{amsthm}
\usepackage{enumerate}

\newtheorem{theorem}{Theorem}

\newtheorem{corollary}{Corollary}

\makeatletter
\let\Im\undefined
 
\DeclareMathOperator{\Im}{Im \,}

\DeclareMathOperator{\tr}{tr}
\DeclareMathOperator{\E}{\mathbb{E}}

\newcommand{\Z}{ {\mathbb Z} }

\newcommand{\G}{\mathbb{G}}
\newcommand{\R}{\mathbb{R}}

\def\Ev#1{{\mathbb E}\left(#1 \right)}

\newcommand{\bra}[1]{\langle #1|} 
\newcommand{\ket}[1]{|#1\rangle}

\def\be{\begin{equation}}
\def\ee{\end{equation}}

\makeatother

\title{Absence of mobility edge for the Anderson random potential on tree graphs 
at weak  disorder}
\shorttitle{Absence of mobility edge for the Anderson model} 

\author{Michael Aizenman\inst{1} \and Simone Warzel\inst{2}}
\shortauthor{M. Aizenman, S. Warzel}

\institute{                    
  \inst{1} Departments of Physics and Mathematics, Princeton University,  Princeton NJ 08544, USA\\
  \inst{2} Zentrum Mathematik, TU M\"unchen,  Boltzmannstrasse 3, 85747 Garching, Germany
}

\abstract{Our recently established criterion for the formation of extended states on tree graphs  in the presence of disorder is shown to have the surprising implication that for bounded random potentials, as in the Anderson model, there is no transition to a spectral regime of Anderson localization, in the form usually envisioned, unless the disorder is strong enough.}

\begin{document}

\maketitle

\section{Introduction}
\begin{figure}
\begin{center}
\includegraphics[width=0.3\textwidth]{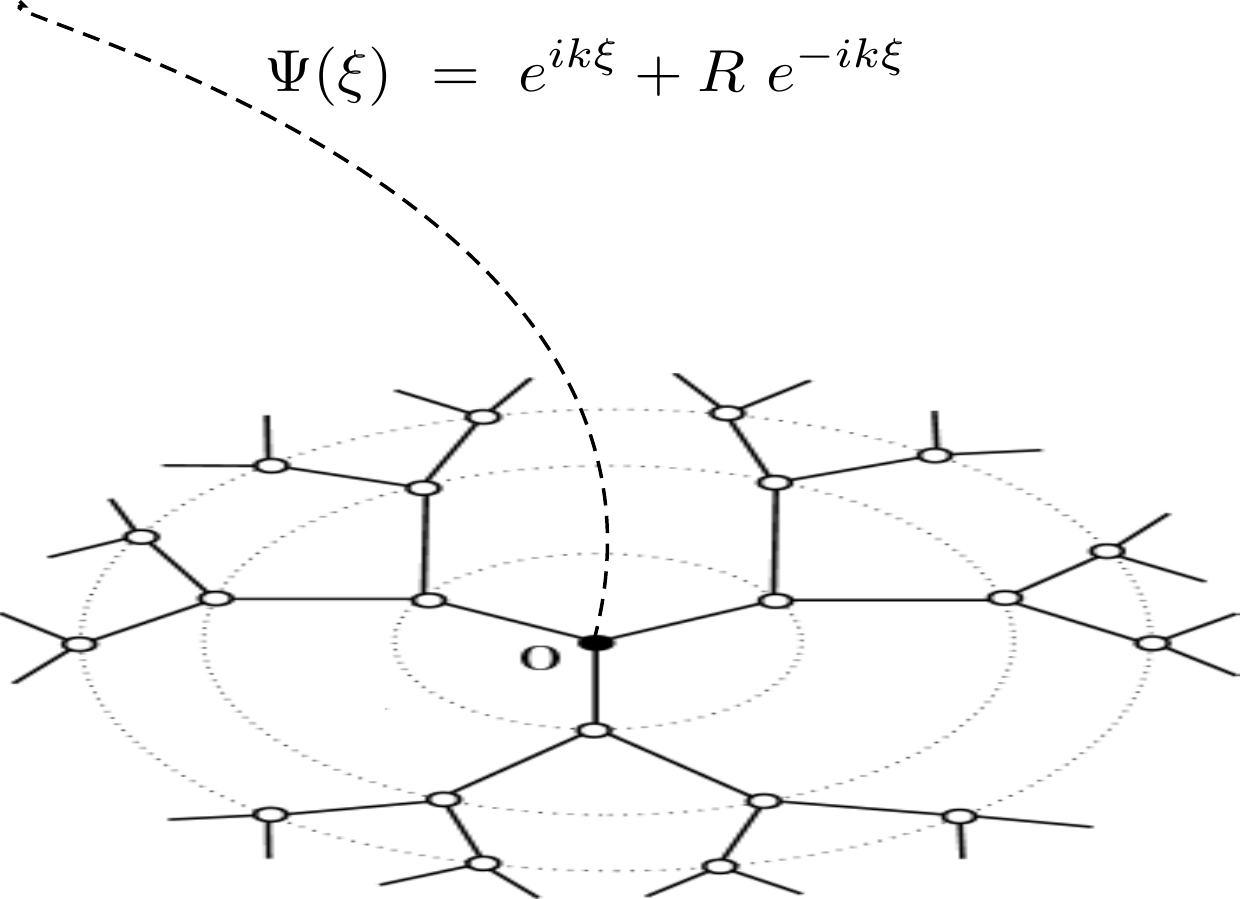}
\caption{A model setup for quantum conduction through the graph (after \cite{MD}): particles are sent  at energy $ E=k^2 +U_{\rm wire} $ down  a wire which is attached to the graph at $ x=0 $.   
The reflection coefficient is strictly less than one, $ |R_E|^2 < 1 $, exactly if~\eqref{eq:ac} holds, which is also the condition for the presence of an absolutely continuous component in the spectrum of 
$H_\lambda(\omega)$.}
\label{fig:wire} 
\end{center}
\end{figure}

Since the early studies of Anderson localization, random operators on tree graphs have provided a testing ground for insights on the effect of disorder
on quantum spectra and dynamics~\cite{A,AAT,AT}.   The basic phenomenon to which this refers  is the transition in the spectra of Schr\"odinger-type operators with random potential from  regimes of energies with only localized states, to energies with extended states which facilitate conduction, e.g.\ in the sense depicted in Figure~\ref{fig:wire}.   

   Our purpose here is to report on rigorous results which challenge some of the generally shared picture of the mobility edge in such systems on the Bethe lattice, i.e. within a regular tree graph (whose degree here is $K+1$).   These results are not expected to affect the picture of Anderson localization for one-particle Hamiltonians in finite dimensions.  However,  in view of analogies which were drawn between tree graphs and  many-particle configuration spaces~\cite{AGKL} (cf.~\cite{BAA,PH}) the mechanism which causes the surprising effect  may be worth paying attention to.  Its root cause is the formation of extended states through fluctuation-enabled resonances between states which initially (i.e., up to a certain scale) may appear to be localized.

    As was explained in~\cite{AW_PRL2011}, the exponential increase in the volume of the relevant configuration space 
    implies that   
  resonances whose likelihood decays exponentially in the distance 
   will nevertheless occur, and predictably so, provided the rate is small enough.   
Somewhat heuristically: states which may locally appear to be localized have  arbitrarily close energy gaps ($\Delta E$) with other states  to which the  tunneling amplitudes decay  exponentially in the distance $R$, as $ e^{-L_\lambda(E) R} $.   
Mixing between two levels would occur if $\Delta E \ll  e^{-L_\lambda(E) R} $.  The latter is a rather stringent condition, however since the  volume grows exponentially fast (as $K^R$)  extended states may form.  A sufficient condition for that is: $ L_\lambda(E)  < \log K $ (for  the exact condition one needs to take into account also the effects of large deviations).   

As a consequence, even under circumstances  which locally seem to imply localization, 
   e.g. as in the Lifshitz tail regime of low density of states, we find 
extended states.    Some basic rigorous results related to  this mechanism were presented the recent work~\cite{AW_math2011}.  Our focus here is on the somewhat surprising implication for bounded random potentials, as featured in a class of basic models of the Anderson transition. 
 
\section{The basic model} 
A specific example of a  one particle Hamiltonian to which our discussion relates is
\be \label{eq:a+a}
{\mathcal H}(\omega)  \ = \  t \sum_{x \sim y} a^\dagger_x a_y  +  \sum_{x\in {\mathcal G}} \varepsilon_x(\omega)  \, a^\dagger_x a_x 
\ee 
where the first sum is over pairs of neighboring sites of a regular tree graph ${\mathcal G}$, i.e., the Bethe lattice,  and $a^\dagger, a$ are fermionic creation and annihilation operators. 
The disorder is expressed in the random `onsite energies' $\varepsilon_x(\omega)$ which are given by independent identically distributed (\emph{iid}) random variables, e.g., of a uniform distribution in the interval $[-W/2,W/2]$.  A  more general class is described below.

In the one-particle sector, the above Hamiltonian 
may also be written, as is often done in the mathematical literature on the subject, 
as
\be \label{eq:H}
H_\lambda(\omega) \ = \ T  +\lambda\,  V(\omega) \quad \mbox{on $ \ell^2(\mathcal{G}) $,}
\ee 
where: $T$ is the graph's adjacency operator, which acts as $(T\psi)(x) = \sum_{y\sim x}\psi(y)$,   $V(\omega)$ is a random potential, and $\omega$ represents the randomness.    The strength of the disorder is controlled by the parameter $\lambda~(\ge0)$, which in the  notation of \eqref{eq:a+a} corresponds to $W/(2t)$.

The results presented below are formulated for random potentials whose values at the graph sites, $ x \in \mathcal{G} $,  form \emph{iid} random variables $V_x(\omega)$  whose probability distribution is of the form $\rho(V) dV$   with a density function $\rho(V) $  which satisfies a simply stated regularity condition, which is spelled below.  
Our emphasis here is on bounded random potentials,  and more specifically on the 
 case when  the support of $\rho$,
which is the closure of the set on which $\rho(V) \neq 0$, is 
exactly $[-1,1]$.    A simple example to which the discussion applies is the `Anderson random potential' with $\rho(V)$ constant over the interval $[-1,1]$ and zero elsewhere. 

On homogenous graphs, such as the regular lattice $\Z^d$ or the homogenous trees discussed here,  the spectrum   itself is easy to determine for random operators such as $H_\lambda(\omega)$:   by ergodicity arguments~~\cite{Pa80,KS_cmp,CFKS} it is the set sum of the spectrum of $T$ and of $\lambda V$.   More explicitly,  for tree graphs with  branching number $K $ (both the rooted and the homogeneous case) and potentials described above,  the spectrum of $H_\lambda(\omega)$ is for almost all realizations given by the non-random set
\begin{eqnarray}    \label{eq:spec}
\sigma(H_\lambda) =  
[-|E_\lambda|, |E_\lambda|] \, , \quad \mbox{with $E_\lambda =  -(2\sqrt{K} + \lambda)$}
\end{eqnarray} 
 (An auxiliary result of \cite{JL01} clarifies that the spectral measures associated to the different vectors $\delta_x\in \ell^2(\G)$ are almost surely all equivalent.)   
Thus, as the strength of the disorder is increased from $\lambda = 0$ upward the outer edges of the spectrum spread at the linear rate $\lambda$.   However, the nature of the spectrum is somewhat less obvious.

\section{The phase diagram} 
\begin{figure} 
\begin{center}
\includegraphics[width=0.45\textwidth]{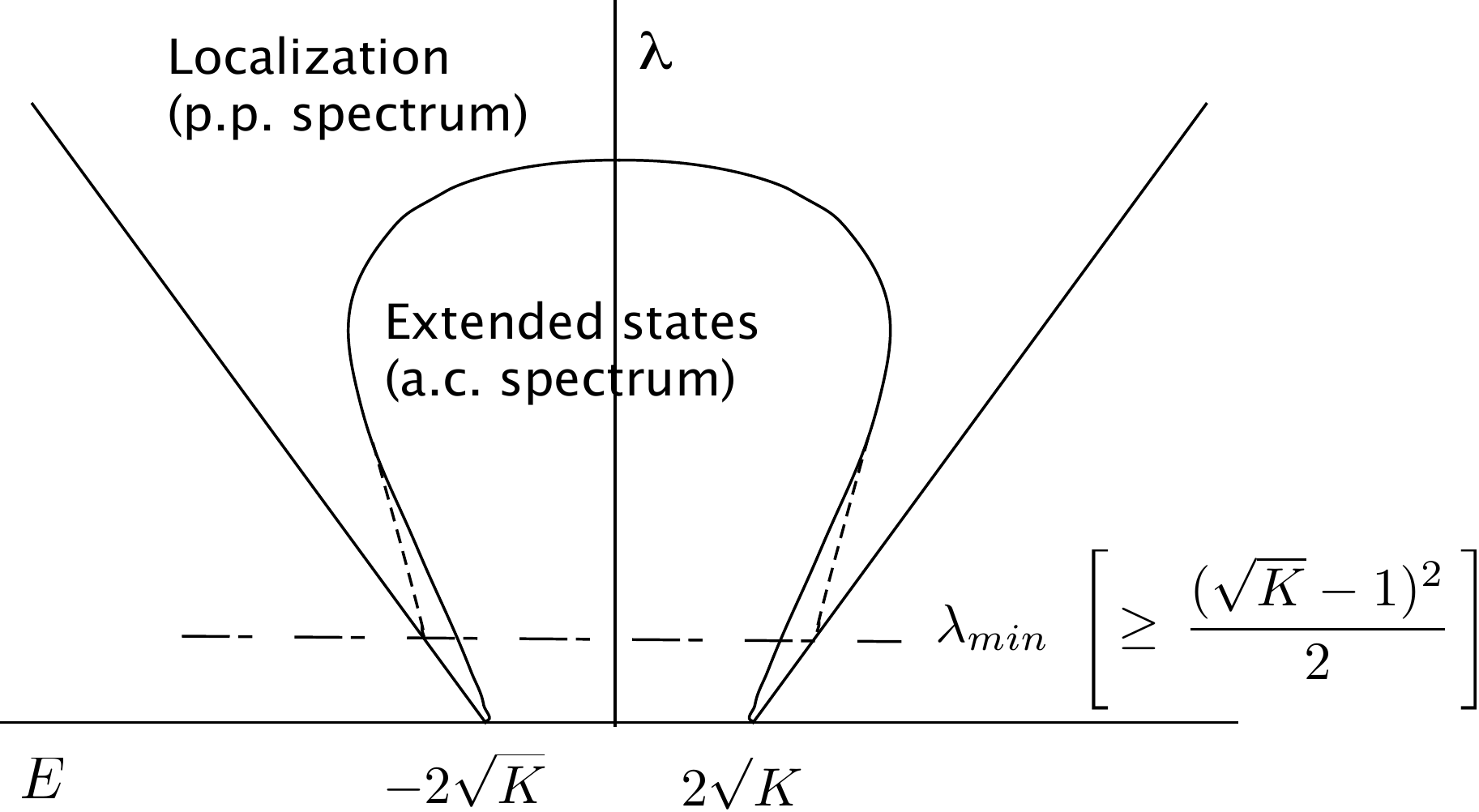}  
\caption{A schematic depiction of the previously expected phase diagram for the Anderson model on the Bethe lattice (the solid line) and the correction presented here (dashed line).  The previous incomplete theoretical analysis~\cite{AAT,AT} has received some support from  extensive numerical work \cite{BST}.  Our analysis suggests that at weak disorder  there is no localization and the spectrum is purely \emph{ac}.  While the proof of that is incomplete, we do prove that for $\lambda \le  (\sqrt{K}-1)^2/2 )$
there is no mobility edge beyond which localization sets in.  }
\label{numMobEdge} 
\end{center}
\end{figure}

It is known that in one dimension complete localization sets in even at arbitrarily weak disorder, with the localization length being finite (except at isolated energies in some special  cases~\cite{LGP,SimSto}) 
and dependent on the disorder strength~\cite{MT,GMP}.  However, it is expected~\cite{AALR} that above two dimensions  at weak and intermediate disorder the random operator exhibits both spectral regimes, with a phase diagram, which in case ${\rm supp}~\rho = [-1,1]  $ is roughly outlined in Fig.~\ref{numMobEdge}.

The above picture  has also been expected to describe the phase diagram for the Bethe lattice.  That is: 
\begin{enumerate}[i.]
\item At weak and moderate disorder  a mobility edge has been expected, beyond which 
the spectrum consists of a dense countable collection of  eigenvalues with localized eigenfunctions.  In the central regime the spectrum is absolutely continuous (a property henceforth referred to as \emph{ac}) with extended states which play an important role for conduction. 
\item The extended states disappear at strong enough disorder ($\lambda > \lambda_{\rm sd}(K)$), where  complete localization prevails.  
\end{enumerate} 
Our results show that for  the Bethe lattice this  picture needs to be modified.  

Significant parts of this picture 
are supported by rigorous results, in particular at strong disorder~\cite{AM,A_wd,AHSF,AW_PRL2011} 
(though some questions remain as to the precise asymptotics of $\lambda_{\rm sd}(K)$ for $K\to \infty$). 
The existence of \emph{ac} spectrum for tree graphs was also established~\cite{K,ASW,FHS,AW_PRL2011,AW_math2011}.

Numerically, both \eqref{eq:spec} and the above point {\it i.\/}   have been hard to see in exact diagonalization of finite systems, and/or in simulations of the infinite Bethe lattice.  
Researchers who carried such work report that Eq.~\eqref{eq:spec}  (whose proof is undisputed, and not complicated) 
 is hard to reconcile with numerical evidence~\cite{MD,BST}.  The reason is easy to understand:  
  since at the spectral fringes the mean density of states is very low,  large simulations are needed in order to find evidence of the existence of eigenstates there.    
Naturally, it should be even harder to  probe reliably the nature of states with energies in that regime.  These eigenstates are associated with rare occurrences of nearly extremal values of the potential, realized over sufficiently large connected sets.  Since such  events occur only in well separated locations, the states may initially appear to be localized.  It is perhaps because of that, as well as the general theoretical guidance offered by the earlier works on localization, that some  numerical studies have been interpreted as providing evidence for the existence of a mobility edge which connects to the unperturbed spectrum $\sigma(T)$, as 
sketched  by the solid line in  Figure~\ref{numMobEdge}. 
 
 The general picture of the phase diagram for unbounded potentials was only recently resolved.  As described in Refs.~\cite{AW_PRL2011,AW_math2011}, on tree graphs  the \emph{ac}  spectrum is fairly robust, and it even appears under arbitrarily weak disorder at energies which are 
remote from that of the unperturbed spectrum  $\sigma(T) = [-2\sqrt{K},2\sqrt{K}]$.   The mechanism which was identified there is also behind the more surprising results for  bounded potentials to which we turn next.

\section{Our main result}  
It is helpful to first introduce the following quantity, which (for reasons related to a dynamical system perspective) is referred to as the Lyapunov exponent:
\be \label{eq:Lyap}
L_\lambda (E) \ :=  \  - \lim_{\eta \searrow 0} \Ev{\log \left |\bra{ 0}  \frac{1}{H_\lambda - (E+i \eta)} \ket{0} \right | } \, ,  
\ee 
where the expectation value sign $ \Ev{\cdot} $ stands for average over the disorder. As is explained in greater detail in \cite{AW_math2011}, on tree graphs the Lyapunov exponent provides the \emph{typical} decay rate of the Green function:
\be 
      |\bra{ 0}  \frac{1}{H_\lambda - (E+i0)} \ket{x}  | \ \approx \  \text{Const.} \, e^{-L_\lambda(E) |x|} 
\ee 
with $(E+i 0)$ indicating the limit $\eta\searrow 0$, as in \eqref{eq:Lyap}, which exists for almost all energies.

The results mentioned below are derived under a regularity conditions for the probability distribution of the potential.   In its general form (a simpler one is presented below) it is stated in terms of the ``minimal function'', which for an integrable function $g$ on $\R$, we define as:  
$M_g(V)  := \inf_{\nu  \in (0, 1]}  \int_{|u-V|\le   \nu}  g(u)  du/(2\nu)  $. \\

\noindent{\bf Definition}  { \it  We call probability distribution on the line  {\it M}-regular if it is absolutely continuous and its density function is bounded relative to its minimal function, i.e., 
\be  \label{def:max}
\rho(V) \ \le \  b \, M_\rho(V)\, ,    \quad \mbox{for almost every $V\in \R$} \,.  
\ee   
   at some  $b<\infty$. }\\  
A simple sufficiency condition for \eqref{def:max} is that $\rho$ is bounded and there is a partition of the line into a finite collection of intervals within each of which $\rho$ is continuous and monotone, except possibly at the interval's boundary  (c.f.\cite{AW_math2011}).  

The  results presented here start from the following  extension of the main finding  which was reported in \cite{AW_PRL2011}, and which was now shown to be valid also for bounded random potentials~\cite{AW_math2011}.
 
\begin{theorem}  \label{thm:L}   Let $V$ be a random potential whose values are given by independent identically distributed random variables with an  {\it M}-regular probability distribution.  Then, for the operator  defined by \eqref{eq:H} on a $K$-regular tree,  at 
almost every energy $E\in \sigma(H_\lambda)  $ at which also 
\be \label{eq:Lyap_cond}
  L_\lambda(E) < \log K    \, ,   
\ee 
one has, with probability one:  
\be \label{eq:ac}
 \Im \bra{ x}  \frac{1}{H_\lambda - (E+i0)} \ket{x} \  > \ 0 \, ,    \quad \mbox{ at any $x\in \mathcal{G}$.    }
\ee 
\end{theorem} 
The dynamical significance of~\eqref{eq:ac} 
was nicely illustrated in~\cite{MD}: if quantum particles are sent coherently at energy $E=k^2+U_{wire}$ down a wire attached to the graph at the vertex $x$, the reflection coefficient $R_E$ satisfies $|R_E|< 1$ if and only if \eqref{eq:ac} holds  (cf.~Fig.~\ref{fig:wire} and \cite{ASWb}).   
%
%
%
More explicitly, in the wire the particles' state is described by plane waves, the wave function at position $ \xi $ being 
$ e^{ik_E\xi} + R_E \, e^{-ik_E\xi} $.  The stationary state is given by a solution of the equation
$(H-E) \Psi =0$ in $\ell^2({\mathcal{G}}) \oplus L^2(\mathcal{Z}_+)$ (with ${\mathcal{G}}$ the graph and $\mathcal{Z}_+$ the wire).   The matching conditions for this equation at $x$ relate the reflection coefficient $ R_E $ to the Green function's   value at $ (x,x) $,  and yield the equivalence:
\be \label{eq:R}
 |R_E| < 1 \quad\mbox{if and only if \eqref{eq:ac} is satisfied.}  
 \ee

 The condition \eqref{eq:ac} is also of direct spectral significance:  
the quantity $ \pi^{-1}  \Im \bra{ x}  \frac{1}{H_\lambda - (E+i0)} \ket{x}  $ is the density of the \emph{ac} component of the spectral measure associated with the state $ \ket{x} $.  
Thus the criterion of Theorem~\ref{thm:L} readily implies  
that in  any energy interval  within  $\sigma(H_\lambda(\omega)) $ in which \eqref{eq:ac} holds on a set of energies of positive Lebesgue measure, the random operator almost surely  has some \emph{ac} spectrum (and hence also extended states, at the corresponding energies).

It is  therefore of relevance to learn about the region in the phase diagram in which the condition~\eqref{eq:Lyap_cond} holds.  
  In the absence of disorder, i.e. at $\lambda =0$  the Lyapunov exponent is easy to compute, and one finds:  
  \be \label{eq:L0}
		  L_0(E) \  \left\{ \begin{array}{ll}
		  	= \log \sqrt{K} \quad &  |E | \leq 2 \sqrt{K} \, , \\  
		 \in \left(\log \sqrt{K} , \log K \right) \quad &  2 \sqrt{K} < |E| < K+1\, ,  \\
		  \geq \log K \quad & |E| \geq K+1 \, . 
		 \end{array} \right.
	\ee

Notably, the  range of energies at which the condition~\eqref{eq:Lyap_cond}  holds at $\lambda=0$ (that is  $ |E| <  K+1 $), 
 is strictly larger   than the spectrum of $H_0$ (which is $ |E| <  2\sqrt{K}$).  
In the context of unbounded potentials  (for which 
the spectrum $ \sigma(H_\lambda(\omega) ) $  covers the whole line as soon as $\lambda\neq0$), employing continuity arguments we  previously showed that  Theorem~\ref{thm:L} carries the following implication~\cite{AW_PRL2011,AW_math2011}. 
\begin{corollary}\label{cor1}
Under the assumptions in Theorem~\ref{thm:L},  for an unbounded random potential, with  ${\rm supp}~\rho = \mathbb{R} $, for any interval $ I\subset (-(K+1), K+1) $ at sufficiently small disorder the random operator has  \emph{ac} spectrum within $I$.  
 \end{corollary}

 For bounded random potentials 
 the spectrum of $H_\lambda(\omega)$ expands at  a bounded rate as $\lambda$ is increased from $0$.   In that situation   
 the above continuity statement does not yield much  information about the spectrum outside of $\sigma(T)$.   For the same reason, the general result of~\cite{A_wd} which in essence proves localization at weak disorder for energies at which $\bra{0} (H_0-E)^{-1}) \ket {0}$ is summable, is also not applicable.  
 To apply Theorem 1 to this case, we present here the following result. 
 
\begin{theorem} \label{thm:ub}
Let $H_\lambda(\omega)$ be a random operator of the form \eqref{eq:H} with a bounded random potential taking values in  $[-1,1] $.  Then for all $\lambda$ small enough, and in particular: 
  \be \label{eq:smallLambda} 
  \lambda < \Delta_K :=  (\sqrt{K}-1)^2/2\, ,   
\ee 
the condition \eqref{eq:Lyap_cond} holds for all energies satisfying 
\be \label{eq:dr}
|E| > |E_\lambda| - \delta( \lambda) 
\ee 
at some $\delta( \lambda) >0$. 
\end{theorem} 
One may note that the regime defined by \eqref{eq:dr} includes the  spectral edges   
\be  \label{eq:spedge}
[\, E_\lambda, E_\lambda \  + \   \delta( \lambda)\, ) \, \cup \,  (\, |E_\lambda| \  - \   \delta( \lambda), |E_\lambda|\, ]   \, 
\ee 
 as well  the complement of the spectrum.  The current proof yields only a rapidly vanishing value for 
 $\delta(\lambda)$.  However, as mentioned in the discussion below,  the result may be indicative of a much stronger statement.   

This leads us to the following somewhat surprising conclusion. 
   \begin{corollary}\label{cor2}
Let $H_\lambda(\omega)$ be a random operator of the form \eqref{eq:H} with a bounded random potential whose density function $\rho(V)$ satisfies the regularity assumptions of Theorem~\ref{thm:L} and has is supported in $ {\rm supp}~\rho = [-1,1] $ .  
Then, for    $\lambda $ small enough, as in Theorem~2, 
near the spectrum's boundary (i.e., in the spectral edges described by \eqref{eq:spedge}) 
the random operator  almost surely has only purely absolutely continuous spectrum.  
\end{corollary}
The proof is by a direct application of Theorems 1 and 2, combined with the criterion of \cite{Aronszajn,SimWolff}, to which our attention was called by Mira Shamis.  The latter  states that  for operators with random potential a sufficient condition for the spectral measure to be purely absolutely continuous in a given interval is that for almost every energy there condition \eqref{eq:ac} holds almost surely.    

Thus, for the bounded random potentials discussed here, at weak disorder there is no mobility edge beyond which the spectrum is pure point and localization sets in as has been expected (cf.~Fig.~\ref{numMobEdge}).

\section{Outline of the proof of Theorem 2}  

First let us explain the explicit condition \eqref{eq:smallLambda}.  Underlying it is  the following deterministic lower bound, which is valid for any potential satisfying $|V(x)| \le 1$ at all $x$, and $ E_\lambda \equiv -2\sqrt{K} -\lambda$: 
\be \label{eq:lb}
\bra{0} \frac{1}{H_0 + \lambda V - E_\lambda } \ket{0} \  \ge \ 
\bra{0} \frac{1}{H_0  - (E_0  - 2 \lambda)} \ket{0}   \, .  
\ee
This  relation (between positive terms) follows by the monotonicity of the operator-valued function $A \mapsto 1/A$, for positive operators $A$, combined with  the elementary observation that:
 \begin{align}  
 0 \ \le \   H_0 + \lambda V - E_\lambda  \  & \le  \  H_0 - (E_0  - 2 \lambda)\ 
 \end{align}   
 
Hence, for any $\lambda < [(K+1)-2\sqrt{K}]/2 = \Delta_K $:  
 \be 
  L_\lambda(E_\lambda) \  \le \   L_0(E_0-2\lambda)  <  \log K \,. 
  \ee  
  where the second inequality is by   \eqref{eq:L0}.   Consulting the exact values of the Lyapunov exponent, which  is computable for $\lambda =0$ (c.f.~\cite{AW_math2011}), one can say more:
\be  \label{eq:Lub}
\mbox{for any $\lambda < \Delta_K $:}  \quad L_\lambda(E_\lambda) < \log K - C(\Delta_K -\lambda) \, , 
\ee 
at some $C>0$. 

 Thus,  for $\lambda < \Delta_K$ 
the condition~\eqref{eq:Lyap_cond}
continues to hold at the boundary of the spectrum.   To prove Theorem~\ref{thm:ub} we need to  show that it continues to do so also within an interval which extends into the spectrum.    Following is a summary of the main ingredients of  the proof of that.  \\

\noindent {1.\/} To study the resolvent $ G_\lambda (x,y;E) = \bra{x} \frac{1}{H_\lambda - E -i0} \ket{y} $ at energies in the range $E\in [E_\lambda, E_\lambda + \Delta E]$, we compare it to the corresponding quantity for  the restrictions of the operator to  finite   subtrees of depth $R$ (measured from the root).  We denote the corresponding operator  by $H^{(R)}_\lambda(\omega)$, and the matrix elements of its resolvent by 
$ G_\lambda^{(R)} (x,y;E) $.  \\  

 \noindent {2.\/} It is of relevance here to note that  by the variational principle the restriction raises the lower edge of the spectrum. 
 The values of $\Delta E >0$ and $R<\infty $ are selected so that with probability sufficiently close to one  (as determined by the rest of the argument)
 \be \label{eq:deltaE}
 \inf \sigma(H^{(R)}_\lambda(\omega)) \ge    E_\lambda  + \Delta E\, . 
 \ee    
This is done with the help of the explicit probability estimate 
	\begin{equation}\label{eq:LT}
		\mathbb{P}\left(   \inf \sigma(H_\lambda^{(R)} ) < E_\lambda + \Delta E\right)  \ \leq \  C \, K^R \,  (\Delta E)^{3/2} 
	\end{equation}
at some $C<\infty$.   This is based on a familiar argument  for such a purpose, using as input the available bounds on $\E  {\tr }e^{-t H_0}$.
 (The bound seems far from optimal, as it does not take into full account the expected Lifshitz tail decay of the density of states at the edge by which the power law  
$(\Delta E)^{3/2}$  in~\eqref{eq:LT} could be replaced by a much faster decaying  rate, e.g., perhaps as discussed in the recent   work~\cite{BapSem2011_LifT}). \\ 

 \noindent {3.\/}  Proceeding under the assumption that \eqref{eq:deltaE} holds, we show that  the finite volume analog of the Lyapunov exponent, $L_\lambda^{(R)}(E)$, for which $H_\lambda(\omega)$ is replaced in \eqref{eq:Lyap} by 
$H_\lambda^{(R)}(\omega)$, satisfies the analog of \eqref{eq:Lyap_cond} for all energies $E < E_\lambda + \Delta E$.  The argument is facilitated by positivity arguments which are applicable under the above assumption.\\

 \noindent {4.\/}   The difference $|L_\lambda^{(R)}(E) - L_\lambda(E)|$ is to be estimated using bounds on the difference of the corresponding Green functions.  For the latter, the resolvent identity yields
\begin{multline}\label{eq:pertGR}
	\left| G_\lambda(0,0;E) - G_\lambda^{(R)}(0,0;E)  \right| \ \leq \\[2ex] 
	\! \sum_{x :\, |x|=R} \big| G_\lambda^{(R)}(0,x_- ;E) \big| \, \left| G_\lambda(0,x ;E) \right| 
	\ =: \ S_\lambda^{(R)}(E) \, . 
\end{multline}
with $|x|$ the distance to the root, and $x_-$ the site preceding $x$ relative to the root.  \\  

 \noindent {5.\/}  To estimate the error term $S_\lambda^{(R)}(E)$, we prove that in case~\eqref{eq:deltaE} holds 
 $| G_\lambda^{(R)}(0,x_- ;E) \big| $ decays at a faster rate than $1/\sqrt{K}^R$.  
 More explicitly,  
 for all $ \lambda > 0$ there are $\delta(\lambda) >0$, and  $C(\lambda), \widehat C(\lambda,s) <\infty$ such that:  \\  
   {\it i.\/}
  Except for a rare event, of probability $\varepsilon(R,\Delta, \lambda )$
 \be  
 \big| G_\lambda^{(R)}(0,x_- ;E) \big| \ \le \  \frac{C(\lambda)}{K^{(1/2 +\delta(\lambda) ) R} }  \, .
 \ee 
simultaneously for all $x$ with $|x|=R$. \\ 
{\it ii.\/}
 For the other term  the decay of the moments $s\in (0,1)$ is not  slower than at the rate $1/\sqrt{K}^R$, in the sense that 
for $x$ as above 
\be  
\E\left( \big| G_\lambda (0,x;E) \big|^s \right) \ \le \  \frac{\widehat C(\lambda,s)}{K^{s  |x|/2  } }  \, . 
 \ee 
 As a consequence, the event 
 \begin{equation}\label{eq:Ssmall}
 S_\lambda^{(R)}(E) \ \leq \ G_0(0,0;E_0-2\lambda) \, K^{-\delta(\lambda) R/2} 
 \end{equation}
 with a suitably small $ \delta(\lambda) > 0 $ has a probability which decays exponentially in $ R $. \\
 
 \noindent {6.\/}    The above results are combined into the observation that in case~\eqref{eq:deltaE} and~\eqref{eq:Ssmall} hold:
  \begin{align} \label{eq:mainterm}
 & | G_\lambda(0,0;E) | \geq  G_\lambda^{(R)}(0,0;E) - S_\lambda^{(R)}(E) \\
 &  \geq G_0(0,0;E_0-2\lambda) \, ( 1 - K^R e^{2RL_0(E_0-2\lambda)} - K^{-\delta(\lambda) R/2}) \, . \notag
  \end{align}
where the last factor can be made arbitrarily close to $1$ through an appropriate choice of $R$.   
The contribution to the expectation of value of $ \log | G_\lambda(0,0;E) |$, a quantity which is of finite variance, due to the  potential configurations when either~\eqref{eq:deltaE} or~\eqref{eq:Ssmall} fails is controlled through the square root of the probability of that rare event.    Since the main contribution is described by 
\eqref{eq:mainterm}  we find that   
$ L_\lambda(E) $ is bounded from above by $ L_0(E_0-2\lambda) = - \log G_0(0,0;E_0-2\lambda) $ plus a correction which vanishes as $\Delta E \to 0$.  This concludes the proof of Theorem~\ref{thm:ub}.\\

The full details of the analysis outlined above are provided in the lengthier and more technical article \cite{AW_math2011}.   Our goal in the above summary has been to convey the essence of the result, and the idea of the proof.   
Since the focus here was limited to what is presently provable rather than what may be more generally true, we add comments on the latter in the discussion which follows.

\section{Discussion} 
The relevant mechanism behind the  results reported here is the formation of extended states out of local modes which up to a certain distance may appear  to yield localized eigenstates.  These    
may join to form extended states through rare resonances, when the gap between the local self energies is smaller than the tunneling amplitude.  The latter   decays exponentially  in the distance.  Therefore the two  important factors for the effect are the local fluctuations in the self energy, and 
the exponential growth of the volume of the configuration space.  Under the Lyapunov exponent condition \eqref{eq:Lyap_cond} the rare resonance conditions are actually met even at large distances.  The implication for the bounded potential, namely the absence of a mobility edge at weak disorder at the edges of the spectrum which is reported here, is then a natural consequence of the (at least) partial continuity of the Lyapunov exponent.  

Let us add that more may be true:  Guided by the expectation that $L_\lambda(E)$ is jointly continuous in $(\lambda,E)$ throughout the spectrum, we conjecture that for bounded potentials, of sufficiently regular distribution, at weak enough disorder there is no localization at all  and the spectrum is purely absolutely continuous.   Nevertheless, both the generalized eigenstates and the time evolution may given the impression of the existence of localization,  but that should be limited to finite  time and distance scales .  

Another  question which may warrant further attention, is whether the mechanism discussed here plays a similar role in case of particle systems with short range interactions.  There are  profound differences between the two systems, but also there the volume of the set of configuration reachable by $R$ steps (when $R$ is of the order of the volume) grows exponentially fast.

\acknowledgments
We thank Mira Shamis for a useful discussion, and the Weizmann Institute of Science, where some of the work was done, for  hospitality.  The work was supported in part by NSF grants PHY-1104596, DMS-0602360 (MA) and DMS-0701181 (SW).

 \end{document}